\documentclass{article}
\usepackage{spconf,amsmath,graphicx,hyperref}
\usepackage{dblfloatfix}
\usepackage{xcolor}
\usepackage{comment}
\usepackage{pifont}
\newcommand{\cmark}{\ding{51}} 
\newcommand{\xmark}{\ding{55}} 

\usepackage{amsmath, amssymb, amsfonts, mathtools}

\usepackage{newunicodechar}
\usepackage[ruled,vlined,linesnumbered]{algorithm2e}
\usepackage{float} 
\usepackage{booktabs} 
\usepackage{amssymb}  
\usepackage{array}    

\newcommand{\blanktok}{\texttt{\textless blank\textgreater}}

\usepackage{booktabs,makecell,array,pifont}

\usepackage{amsmath,amssymb,amsfonts,mathtools}
\usepackage{makecell}
\usepackage{multirow}
\usepackage{booktabs}
\usepackage{hyperref}
\usepackage{array}
\usepackage{times}     

\newcolumntype{L}[1]{>{\raggedright\arraybackslash}p{#1}}

\renewcommand{\cmark}{\checkmark}
\renewcommand{\xmark}{$\times$}
\ninept

\usepackage{etoolbox}

\title{TASU: Text-Only Alignment for Speech Understanding}
\name{Jing Peng$^{1}$, Yi Yang$^{1}$, Xu Li$^4$, Yu Xi$^{1}$, Quanwei Tang$^{3,4}$, Yangui Fang$^{2,4}$, Junjie Li$^{1}$, Kai Yu$^{1}$$^{\dagger}$\thanks {$^{\dagger}$Corresponding author.\\  \protect\href{https://github.com/PigeonDan1/ps-slm.git}{\underline{https://github.com/PigeonDan1/ps-slm.git}}}
}

\address{
\normalsize
$^1$X-LANCE Lab, School of Computer Science, Shanghai Jiao Tong University, Shanghai, China\\
\normalsize
$^1$MoE Key Lab of Artificial Intelligence,  $^1$Jiangsu Key Lab of Language Computing\\
\normalsize
$^2$School of Electronic Information and Communications, Huazhong University of Science and Technology, China \\ 
\normalsize
$^3$School of Computer Science \& Technology, NLP Lab, Soochow University, China \quad $^4$AISpeech Ltd, Suzhou, China\\
\normalsize
\{jing.peng, yuxi.cs, junjieli, kai.yu\}@sjtu.edu.cn~~~fangyg@hust.edu.cn~~~ qwtang101@stu.suda.edu.cn~~~ xu.li@aispeech.com~~~ 
}
\begin{document}
%
\maketitle

\begin{abstract}

  Recent advances in Speech Large Language Models (Speech LLMs) have paved the way for unified architectures across diverse speech understanding tasks. However, prevailing alignment paradigms rely heavily on large-scale audio-text paired data and computationally intensive training, yet often exhibit limited generalization to unseen domains or tasks. To address these limitations, we propose TASU (Text-only Alignment for Speech Understanding), a novel alignment paradigm that can leverage only unpaired text data to guide cross-modal alignment. Experiments show that TASU achieves competitive zero-shot speech recognition. Leveraging this property, it can further function as a pre-training stage in curriculum learning, enhancing domain generalization in speech recognition. Ultimately, TASU can extend its zero-shot generalization to a wide range of speech understanding tasks and notably outperforms prominent Speech LLMs including \textit{GLM-4-Voice} and \textit{Step-Audio} on the \textit{MMSU} benchmark, establishing TASU as an efficient and scalable alignment paradigm for Speech LLMs.
  
\end{abstract}


\section{Introduction}
\label{sec:intro}

In recent years, large language models (LLMs) have demonstrated remarkable capability in contextual reasoning and multitask learning, and have been increasingly applied to speech understanding~\cite{arora2025landscape,peng2024survey}. Unlike traditional cascaded systems that rely on automatic speech recognition (ASR) to provide textual input, modern Speech LLMs align speech and text modalities directly through mechanisms such as continuous feature projection or discrete token augmentation~\cite{chu2024qwen2,tang2023salmonn,ma2024embarrassingly,zhang2023speechgpt}. These approaches have enabled state-of-the-art (SOTA) performance in both single-task settings and broad multi-task speech understanding~\cite{xu2025fireredasr,bai2024seed,geng2025osum, wang2022optimizing}.  

However, existing alignment paradigms face two major limitations. 
First, continuous feature projection, though capable of preserving detailed audio information, often introduces substantial redundancy. 
Such redundancy not only increases computational cost during training and inference but also raises the risk of overfitting~\cite{fang2025low}. Second, mitigating these issues typically requires massive amounts of paired audio--text data and complex training pipelines in order to achieve competitive multitask performance~\cite{xu2025qwen2,ding2025kimi}.  

To alleviate the issue of redundancy in continuous audio features, earlier studies explored alternative representation refinement techniques. The \emph{CTC lattice}, first introduced in~\cite{chen2016phone}, organizes frame-level CTC posterior distributions into a compact structure that represents all possible alignment paths. Building on this lattice, Chen \emph{et al.} proposed the Phoneme Synchronous Decoding (PSD) and Label Synchronous Decoding (LSD) methods~\cite{chen2016ctc,chen2016phone,Chen2019LSD}, which exploit CTC~\cite{graves2006connectionist} posteriors to perform efficient variable frame rate search and effectively reduce redundant acoustic frames.
Liu \emph{et al.} further proposed PSD joint training within end-to-end ASR models~\cite{9141380}, verifying that the extracted audio semantic representations accelerate model training with almost no loss of semantic information. This representation also enables the speech and text modalities to be aligned at a comparable level of information flow. 
Moreover, compared with raw audio hidden embeddings, CTC posteriors exhibit stronger structural similarity to text, which makes it possible to approximate them using one-hot vectors derived from transcripts. 
This insight suggests that training can rely on minimal, or even no, real speech data, substantially mitigating the two limitations discussed earlier.

Motivated by these, we propose \textit{TASU (\textbf{T}ext-only \textbf{A}lignment for \textbf{S}peech \textbf{U}nderstanding)}, a novel alignment paradigm that achieves robust cross-modal alignment without relying on audio supervision. We similarly use LSD to extract audio CTC posteriors into compact ``codebook''-like features, preserving semantic content while removing redundancy. From the text side, we introduce \textit{CTC posterior simulation (CPS)}, which mimics real CTC distributions, including frame dropping and repetition, to generate pseudo-``codebooks'' from text-only data. This dual design allows TASU to bridge modalities efficiently while keeping the LLM backbone frozen, thus retaining its inherent multitask capability. In this work, we focus on semantic speech understanding tasks, which are representative of core challenges in spoken language processing and well-suited for evaluating multitask performance in Speech LLMs.

\begin{figure*}[!t]  
  \centering
  \includegraphics[width=0.75\textwidth,trim=0 0 0 0,clip]{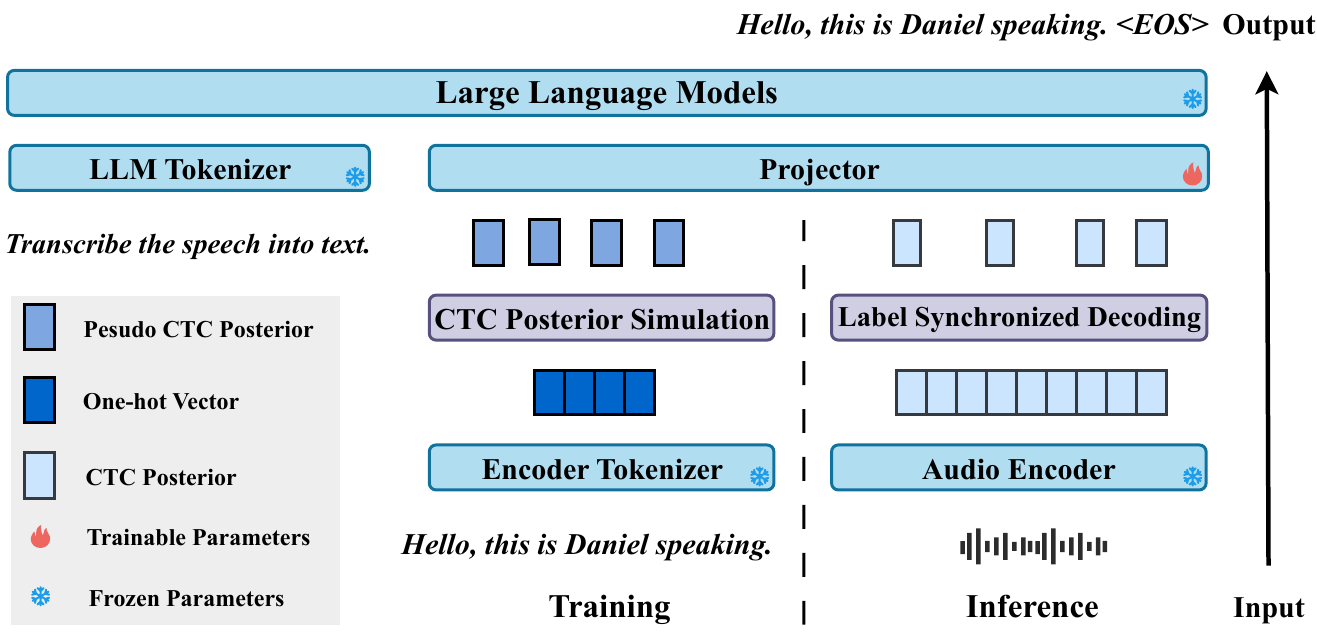}
  \caption{An Overview of TASU: during training (left), only text inputs are used: transcriptions are tokenized into one-hot vectors and converted into pseudo CTC posteriors via simulation. During inference (right), speech is encoded to generate real CTC posteriors, which are refined by label-synchronous decoding. Both pseudo and real CTC posteriors are mapped by a trainable projector into the frozen LLM, producing outputs such as transcriptions or other speech understanding tasks.}
  \label{fig:overview}
\end{figure*}

The main contributions of this work are summarized as follows:

\begin{itemize}
    \item \textbf{Zero-shot Speech Recognition and Domain Generalization:} We show that TASU alone delivers zero-shot ASR with small accuracy degradation relative to audio-text supervision in in-domain evaluation; when leveraged as a curriculum pre-training stage, it further enhances domain generalization while preserving source-domain accuracy.
    
    \item \textbf{Multitask Generalization in Speech Understanding:} TASU enables Speech LLMs to achieve strong zero-shot generalization on speech understanding tasks using limited task-specific text data. On the \textit{MMSU} benchmark~\cite{wang2025mmsumassivemultitaskspoken}, TASU surpasses mainstream alignment paradigms such as SLAM at the same data scale, and further outperforms large-scale speech models including \textit{SALMONN-13B}, \textit{GLM-4-Voice}, and \textit{Step-Audio}.
    
    \item \textbf{TASU: Audio-Efficient and Generalizable Speech–Text Alignment:} LSD achieves nearly 6$\times$ downsampling, greatly accelerating while enhancing semantic extraction and also alleviating overfitting; meanwhile, CPS markedly reduces the reliance on audio data and helps domain generalization in recognition and multitask generalization in speech understanding.

\end{itemize}


\vspace{-1em}
\section{related work}
\vspace{-5pt}
\label{sec: relatedwork}

Connectionist Temporal Classification (CTC) introduces an explicit \textit{blank} symbol and marginalizes over all possible alignments, thereby mapping unsegmented acoustic sequences into variable-length label sequences. Its ``collapse'' operation, which removes blanks and merges repetitions, projects frame-level posteriors into a compact representation that can be directly exploited for decoding~\cite{graves2006connectionist, chen2016ctc}. 

There are already methods that leverage CTC posterior probabilities to address the two issues outlined in the Section~\ref{sec:intro}. First, \textit{AlignFormer} uses CTC signals to downsample acoustic features more effectively and~\cite{fan2025alignformer}, to some extent, demonstrates strong instruction-following ability. However, when relying only on a small amount of paired audio-text data, its multitask performance degrades markedly, with accuracy on multiple-choice tasks approaching chance. In contrast, \textit{LegoSLM} employs CTC posteriors as weights to reweight LLM word embeddings for acoustic representation~\cite{ma2025legoslm}, yielding more structured representations; yet it sacrifices multitask capability and requires large-scale data to re-align the LLM’s vocabulary.

\section{TEXT-ONLY ALIGNMENT FOR SPEECH UNDERSTANDING}
\label{sec:method}

Conventional alignment strategies for Speech LLMs often rely on encoder hidden states with heuristic subsampling, which either produce redundant and noisy representations or risk discarding critical information. In addition, acoustic features exhibit high temporal variability that mismatches the structured nature of text embeddings, making cross-modal alignment challenging. To address these issues, we propose \textbf{TASU}, which aligns speech and text directly at the \emph{CTC posterior level} (Fig.~\ref{fig:overview}). The key idea is to establish a unified posterior interface for both training and inference:  
\begin{itemize}
    \item \textbf{Training:} text transcriptions are tokenized into one-hot vectors and transformed into pseudo-posteriors by the CPS module, which supervise the trainable projector.  
    \item \textbf{Inference:} raw speech is encoded into real CTC posteriors, refined by LSD, and mapped by the pretrained projector into the frozen LLM.  
\end{itemize}

In this way, TASU enables text-only training while ensuring that both modalities share compact, structured, and semantically aligned posterior representations.

\subsection{Label\mbox{-}Synchronous Decoding (LSD)}
CTC decoding typically involves large portions of blank symbols and consecutive repetitions of the same token~\cite{graves2006connectionist,chen2016phone}. 
Directly feeding such posteriors into a Speech LLM would propagate redundancy and obscure semantics. 
LSD is designed to compact the sequence while preserving semantic fidelity through two operations.

\noindent\textbf{(1) Blank-frame removal.}
Given a posterior sequence $\mathbf{P}\!\in\!\mathbb{R}^{T\times V}$ with $T$ frames and vocabulary size $V$, 
frames dominated by blank probability are discarded with a tunable threshold $\tau$:
\begin{equation}
\mathbf{P}'_{t} =
\begin{cases}
\varnothing, & \text{if } P_t(\text{\blanktok}) > \tau,\\[2pt]
\mathbf{P}_{t}, & \text{otherwise.}
\end{cases}
\label{eq:psd-blank}
\end{equation}

\noindent\textbf{(2) Consecutive-frame merging.}
Let $y_t=\arg\max_{k} P_t(k)$ denote the top symbol at frame $t$. 
For each maximum number of consecutive identical frames $S_j$ of identical $y_t$, we average their vectors:
\begin{equation}
\mathbf{P}''_{j} \;=\; \frac{1}{|S_j|}\sum_{t\in S_j}\mathbf{P}'_{t},
\qquad j=1,\dots,J,
\label{eq:psd-merge}
\end{equation}
where $J$ is the number of frames retained after Eq.~\eqref{eq:psd-blank}. 
This process eliminates blank-dominated frames and collapses redundant repetitions, yielding temporally compact posteriors that retain essential information for alignment. The proposed method achieves significant compression of acoustic feature sequences without sacrificing semantic completeness.

\begin{algorithm}[t]
\label{alg: cps}
\caption{CTC Posterior Simulation (CPS)}
\DontPrintSemicolon
\KwIn{Token ID sequence $Y=(y_1, \ldots, y_T)$, vocab size $V$, $(\lambda_{\text{low}}, \lambda_{\text{high}})$, deletion prob $p_{\text{del}}$, insertion ratio $p_{\text{ins}}$, blank id $b$}
\KwOut{Simulated posterior sequence $\tilde{\mathbf{S}}=\{\tilde{\mathbf{p}}_t\}$}
\BlankLine

\tcp{1) Sequence-wise Label Smoothing}
Sample $\alpha \sim \mathcal{U}(\lambda_{\text{low}},\,\lambda_{\text{high}})$\;
Initialize empty sequence $\tilde{\mathbf{S}}$\;
\For{$t=1$ \KwTo $T$}{
  Convert $y_t$ to one-hot vector $\delta_{y_t}$\;
  $\tilde{\mathbf{p}}_t \leftarrow \alpha\delta_{y_t} + (1 - \alpha) \tfrac{1}{V}\mathbf{1}$\;
  Append $\tilde{\mathbf{p}}_t$ to $\tilde{\mathbf{S}}$\;
}
\BlankLine

\tcp{2) Random Deletions}
Create new empty sequence $\tilde{\mathbf{S}}'$\;
\ForEach{$\tilde{\mathbf{p}}_t$ in $\tilde{\mathbf{S}}$}{
  \If{Bernoulli$(1-p_{\text{del}})=1$}{
    Append $\tilde{\mathbf{p}}_t$ to $\tilde{\mathbf{S}}'$\;
  }
}
$\tilde{\mathbf{S}} \leftarrow \tilde{\mathbf{S}}'$\;
\BlankLine

\tcp{3) Random Insertions}
Define $\mathbf{e}_{\text{blank}}$ with $(\mathbf{e}_{\text{blank}})_b=1$\;
$N_{\text{ins}} \leftarrow \lfloor |\tilde{\mathbf{S}}| \times p_{\text{ins}} \rfloor$\;
\For{$i=1$ \KwTo $N_{\text{ins}}$}{
  Choose position $pos$ uniformly in $\{0,\ldots,|\tilde{\mathbf{S}}|\}$\;
  \If{Bernoulli$(0.5)=1$ \textbf{an d} $|\tilde{\mathbf{S}}|>0$}{
    $\mathbf{d} \leftarrow \tilde{\mathbf{S}}[\max(0,pos-1)]$\;
    Insert $\mathbf{d}$ at $pos$ in $\tilde{\mathbf{S}}$\;
  }
  \Else{
    Insert $\mathbf{e}_{\text{blank}}$ at $pos$ in $\tilde{\mathbf{S}}$\;
  }
}
\Return $\tilde{\mathbf{S}}$\;
\end{algorithm}

\subsection{CTC Posterior Simulation (CPS)}

To enable training with text-only data, we propose CPS, which converts each ground-truth token into a pseudo-posterior sequence $\tilde{\mathbf{S}}$. CPS consists of three stochastic stages that mimic the variability of real CTC outputs, as detailed in Algorithm~\ref{alg: cps}:

\noindent\textbf{(1) Random Label Smoothing.}
Given a token $y$, represented as one-hot $\delta_y \in \mathbb{R}^V$, we interpolate it with the uniform distribution to obtain a smoothed posterior:
\begin{equation}
\tilde{\mathbf{p}} = \alpha\delta_y + (1-\alpha)\tfrac{1}{V}\mathbf{1}, \quad
\alpha \sim \mathcal{U}(\lambda_{\text{low}}, \lambda_{\text{high}}).
\label{eq:cps_smoothing}
\end{equation}
This yields the initial sequence $\tilde{\mathbf{S}}=[\tilde{\mathbf{p}}]$.
The random factor $\alpha$ ensures that the generated distributions cover a wide range of confidence levels, resembling the uncertainty of real acoustic posteriors.

\noindent\textbf{(2) Random Deletions.}
Each element of $\tilde{\mathbf{S}}$ is removed independently with probability $p_{\text{del}}$, simulating token drops commonly observed in CTC alignments.
This operation models the fact that non-blank tokens can occasionally disappear due to alignment errors, forcing the system to be robust to missing evidence.

\begin{table}[H]
\centering
\caption{A concise comparison of different alignment strategies for multimodal speech understanding models. Train part: E = encoder, P = projector, L = LLM (parentheses denote optional part).}
\label{tab:alignment_comparison_en}
\renewcommand{\arraystretch}{1.2}
\resizebox{\columnwidth}{!}{
\setlength{\tabcolsep}{5pt}      
\begin{tabular}{l c c c}
\toprule
\textbf{System}  & \textbf{Training Data} & \textbf{Train Part} & \textbf{Zero-shot Multitask} \\
\midrule
SLAM     & (Audio, Text) & P + (L) & \xmark \\

LegoSLM          & (Audio, Text) & (P) + L & \xmark \\

AlignFormer & (Audio, Text) & E + P & \cmark \\

TASU        & \textbf{Text-only} &  P & \cmark \\

\bottomrule
\end{tabular}
}
\end{table}
\vspace{-0.5em}
\noindent\textbf{(3) Random Insertions.}
We perform
\begin{equation}
N_{\text{ins}}=\left\lfloor |\tilde{\mathbf{S}}|\times p_{\text{ins}} \right\rfloor ,
\label{eq:n_ins}
\end{equation}

where $p_{\text{ins}}$ controls the insertion rate. For each insertion, a position $pos \in {0,\dots,|\tilde{\mathbf{S}}|}$ is sampled, and either:
(i) a duplicate of $\tilde{\mathbf{S}}[\max(0,pos-1)]$, or
(ii) a blank one-hot vector $\mathbf{e}_{\text{blank}}$,
is inserted with equal probability.
This step introduces alignment jitter to capture CTC-specific repetitions and blank separations, mitigating CTC imprecision and enhancing robustness, without which performance degrades notably based on experiments.

By combining these three operations, CPS transforms clean symbolic labels into noisy multi-frame pseudo-posteriors that closely approximate the distributional properties of real audio. To provide a more intuitive understanding of how TASU differs from other alignment paradigms, Table~\ref{tab:alignment_comparison_en} provides a concise comparison of alignment paradigms for speech LLMs. In particular, only TASU, trained solely on text, achieves zero-shot performance across multiple tasks with projector parameters trainable only. It is worth noting that LSD achieves an average downsampling ratio of nearly 6 on the experimental data, leading to substantial speedups in both training and inference.

\section{Experimental Details}
\label{sec:experiment}

With the two core processes described in Sec.~\ref{sec:method}, TASU can enable zero-shot transfer from text-only training to speech inference. 
To validate its rationality and effectiveness, we conduct a series of controlled experiments with step-by-step verification. 

\noindent\textbf{Model Architecture.} 
Since TASU relies on reliable CTC posterior probabilities, we employ \textit{SenseVoice-Small} as the speech encoder and \textit{Qwen2.5-1.5B} as the language model backbone. 
The projector is instantiated as a \textit{Linear--SiLU--Linear} module, with only its parameters being trainable. Bottleneck is typically set to 1024. For broader speech understanding tasks, it is set to 2048, as in Table~\ref{tab:speech_understanding}.

\noindent\textbf{Training Data.} 
For ASR, the datasets include \textit{LibriSpeech}, \textit{SlideSpeech}, and \textit{CommonVoice4}.  
For speech-to-text translation (ST), we use \textit{CoVoST2 En$\rightarrow$Zh}, and for spoken instruction understanding, we adopt \textit{SLURP}~\cite{ardila2020common, panayotov2015librispeech,wang2023slidespeechlargescaleslideenrichedaudiovisual,wang2021covost2,bastianelli2020slurp}.  

\noindent\textbf{Training Setup.} 
For LSD, parameter $\tau$ is set to $0.9$. For CPS, we set the label smoothing range $(\lambda_{\text{low}}, \lambda_{\text{high}})$ to $(0.8, 1.0)$, and the deletion and duplication probabilities, $p_{\text{del}}$ and $p_{\text{dup}}$, are both set to $0.05$. The learning rate is fixed at $5\times10^{-5}$, with 5 training epochs. 
Checkpoints are selected when the evaluation loss stops decreasing. 

\noindent\textbf{Evaluation Dataset and Setup.}
We evaluate our model on both ASR and Speech Understanding tasks. For ASR, we report Word Error Rate (WER) on the standard in-domain test sets. To further assess generalization, we employ \textit{TED-LIUM 3}~\cite{hernandez2018tedlium3}, testing robustness to a distinct topical and acoustic domain (lectures). For speech understanding task, we assess performance on the \textit{MMSU} benchmark~\cite{wang2025mmsumassivemultitaskspoken}. WER is computed using the official \textit{Wenet} toolkit~\cite{zhang2022wenet}.

\begin{table}[H] 
    \centering 
    \caption{Comparison of different alignment paradigms. All systems share the same components and training setup with only projector trainable. Libri = LibriSpeech, Ted-3 = TedLium-3, Slide = SlideSpeech. Results are WER\%. TASU (+SFT) denotes a two-stage curriculum learning process.} 
    \label{tab:tasu_curriculum} 
    \setlength{\tabcolsep}{6pt} 
    \renewcommand{\arraystretch}{1.08} 
    \resizebox{\columnwidth}{!}{ 
        \begin{tabular}{l|cc|ccc} 
            \toprule 
            \multirow{2}{*}{System} & \multicolumn{2}{c|}{\textbf{Train Data}} & \multirow{2}{*}{\makecell{Libri \\ \textit{clean/other}}} & \multirow{2}{*}{Slide} & \multirow{2}{*}{Ted-3} \\ 
            \cline{2-3} 
            & Text & (Audio, Text) & & & \\ 
            \hline 
            SLAM & -- & Libri & 3.72 / 8.47 & 18.58 & 20.65 \\ 
            \hline 
            \multirow{2}{*}{TASU} & Libri & -- & 4.57 / 9.90 & 24.07 & 19.36 \\ 
            & Libri + Slide & -- & 4.21 / 10.31 & 18.70 & 13.23 \\ 
            \hline 
            \multirow{2}{*}{TASU (+SFT)} & Libri & Libri & 3.55 / \textbf{7.96} & 17.40 & 14.38 \\ 
            & Libri + Slide & Libri & \textbf{3.06} / 8.04 & \textbf{14.65} & \textbf{11.40} \\ 
            \bottomrule 
        \end{tabular} 
    } 
\end{table}

\vspace{-1.8em}
\section{Results and Evaluation}
\label{sec:evaluation}

In this section, we present experimental results in two parts. First, we show that TASU enables zero-shot speech recognition and, when used as a curriculum pre-training stage, allows models fine-tuned only on source-domain audio data to generalize effectively to new domains.
Second, we evaluate TASU on multitask speech understanding, where it achieves zero-shot generalization from limited text and delivers strong performance on \textit{MMSU} benchmark. 

\vspace{-0.8em}
\subsection{Zero-Shot Recognition and Domain Generalization via TASU Curriculum Pre-training}
\label{sec: curriculum}

To evaluate the effectiveness of TASU in speech recognition, we conduct a series of experiments as summarized in Table~\ref{tab:tasu_curriculum}.  To enable a controlled comparison, we implement the SLAM alignment strategy proposed in \textit{SLAM-LLM} without performing downsampling to avoid potential performance degradation~\cite{ma2024embarrassingly}. On the \textit{LibriSpeech} test-clean and test-other sets, TASU shows only less than 1.5\% WER gap compared to the baseline, demonstrating that it can achieve reasonable semantic alignment without paired audio–text training. Furthermore, when \textit{SlideSpeech} transcripts are incorporated into TASU training, we observe consistent improvements on \textit{SlideSpeech} itself, and even surpass the baseline on \textit{TedLium-3} in new domain.

To further explore scalability, we extend TASU as the pre-training stage of Curriculum Learning. In this stage, \textit{Slidespeech} and \textit{LibriSpeech} text transcripts are used to simulate CTC posteriors for training, followed by fine-tuning with \textit{LibriSpeech} audio–text pairs. Results show that TASU not only maintains performance on the \textit{Librispeech} but also yields substantial gains in both \textit{TedLium-3} and \textit{Slidespeech}. These findings highlight the scalability of TASU in leveraging large-scale text-only resources for domain generalization.

\vspace{0.2em}
\noindent\textbf{Ablation:} To further justify the rationality of the baseline presented in Table~\ref{tab:tasu_curriculum} and LSD, ablation studies were conducted to compare  recognition performance under the current alignment paradigm.
\vspace{-1.8em}
\begin{table}[H]
  \caption{Ablation study on LSD (WER\%). All models are only trained on \textit{Librispeech} with the same structure. CTC refers to CTC posterior. Note that TASU without LSD fails to work, resulting in unusable WER scores.}
  \label{tab:align_single}
  \centering
  \setlength{\tabcolsep}{6pt}
  \renewcommand{\arraystretch}{1.1}
  \resizebox{\columnwidth}{!}{
  \begin{tabular}{l c c | c c c}
    \toprule
     System & Projection Feature & LSD & \makecell{Libri\\ \textit{clean/ other}} & Slide & Ted-3\\
    \hline
     SLAM & Hidden & \xmark & 3.72 / 8.47 & 18.57 & 20.65 \\
     SLAM-CTC & CTC & \xmark & 3.79 / 8.13 & 24.13 & 25.89 \\
     TASU & Pseudo CTC & \xmark & $>100$ & $>100$ & $>100$\\
     \hline 
     SLAM-CTC & CTC & \cmark & \textbf{3.13} / 8.59 & 18.59 & 14.61 \\
     TASU & Pseudo CTC & \cmark & 4.57 / 9.90 & 24.07 & 19.36 \\
     TASU (+SFT) & (Pseudo) CTC & \cmark & 3.55 / \textbf{7.96} & \textbf{17.40} & \textbf{14.38} \\
    \bottomrule
  \end{tabular}
  }
\end{table}

\begin{table}[H]
\centering
\caption{Speech understanding multitask generalization with TASU. 
The models in the upper block are built upon the same components and training setup, and share the same multitask data, while TASU only uses text.
Results are reported as WER\%, BLEU and Accuracy.}

\label{tab:speech_understanding}
\setlength{\tabcolsep}{3pt}
\renewcommand{\arraystretch}{1.05}
\resizebox{\columnwidth}{!}{
\begin{tabular}{l c c | c c c}
\toprule
Model  & Model Size & \makecell{Train Audio \\ Duration (h)} &
\multicolumn{1}{c}{\makecell{LibriSpeech\\ \textit{clean / other}\\(WER\%$\downarrow$)}} &
\multicolumn{1}{c}{\makecell{CoVoST2\\ En$\rightarrow$Zh\\(BLEU$\uparrow$)}} &
\multicolumn{1}{c}{\makecell{MMSU\\(ACC$\uparrow$)}} \\
\hline
TASU    & 1.5B   & \textbf{0}  & 6.47 / 10.35 & 33.35 & 40.32 \\[-2pt]
TASU (+SFT)    & 1.5B   & 0.9k   & 3.28 / 6.91 & 36.51 & 40.48 \\[-2pt]
SLAM & 1.5B & 1.8k  & 3.30 / 7.24 & 37.34 & 36.70 \\[-1pt]
\hline

SALMONN & 13B & $>100$k  & 2.10 / 4.90 & 34.40 & 25.84 \\
GLM-4-Voice & 9B  & $>100$k  & 2.82 / 7.66 & -     & 35.51 \\
Step-Audio & 130B  & $>100$k  & 2.36 / 6.32 & -     & 37.42 \\
Qwen2.5-Omni & 7B & $>100$k  & 2.37 / 4.21 & 41.40 & 60.57 \\
\bottomrule
\end{tabular}
}
\end{table}
\vspace{-0.5em}

\noindent Model architecture and training setup kept unchanged in Table \ref{tab:align_single}. SLAM refers to the alignment paradigm adopted in \textit{SLAM-LLM}. Considering the differences in training configurations and convergence issues of the alignment paradigms, results for \textit{LegoSLM} and \textit{AlignFormer} are not reported. We find that LSD can almost fully preserve the semantic information of speech and also alleviates model overfitting, while playing an indispensable role within TASU. 

\subsection{Speech Understanding Multitask Generalization with TASU}

To further investigate the performance of TASU on multi-task speech semantic understanding, we conduct the experiments summarized in Table~\ref{tab:speech_understanding}. We still consider SLAM method as the baseline: using hidden states as projection features without downsampling, which reflects the prevalent alignment paradigm in most existing Speech LLMs. Given that the SLAM architecture fails to develop multitask capabilities when trained on limited task-specific data, we expanded the training data to ensure a fair comparison: \textit{LibriSpeech} and \textit{CommonVoice4} for ASR, \textit{CoVoST2 En$\rightarrow$Zh} for ST, and \textit{SLURP} for instruction understanding. TASU only uses text, while TASU (+SFT) uses half of audio-text pairs for the second-stage SFT. In addition, to provide a more intuitive assessment of TASU, we further compare it with the results of other Speech LLMs in \textit{MMSU} benchmark~\cite{wang2025mmsumassivemultitaskspoken}.

As we can see, TASU demonstrates strong zero-shot multitask generalization for speech understanding: without any audio–text pairs, it achieves better result on \textit{MMSU} than SLAM. When half of audio-text data is incorporated for SFT, the model shows rapid improvement on the ASR and ST tasks.  Notably, TASU even surpasses several large-scale Speech LLMs, underscoring its efficiency as a lightweight yet effective paradigm for speech understanding.

\vspace{-0.3em}
\section{conclusion and future work}
\label{sec: conclusion}

In this work, we propose TASU, a novel alignment paradigm for Speech LLMs trained solely on text data. On the one hand, TASU enables zero-shot speech recognition with only a minor accuracy drop. It can further serve as the first stage of curriculum learning in ASR, improving performance on new target domains while preserving recognition accuracy on the source domain. On the other hand, TASU delivers strong zero-shot multitask speech understanding with limited text data, highlighting its potential as a simple yet effective paradigm for scalable and generalizable Speech LLMs.

In the future, we aim to further refine the CPS approach to narrow the gap between real CTC posteriors derived from audio and pseudo-posteriors generated from text. This will enable a more accurate audio-free alignment paradigm. Moreover, by incorporating large-scale text data, we plan to explore the scalability and performance of this alignment method on a greater scale.
\clearpage

\section{Acknowledgement}
This work was primarily supported by the National Natural Science Foundation of China (NSFC) under Grant No. 92370206, with additional support from the Yangtze River Delta Science and Technology Innovation Community Joint Research Project under Grant 2024CSJGG01100.

\bibliographystyle{IEEEbib}
\footnotesize
\bibliography{strings,refs}

\end{document}